\documentstyle[preprint,pra,aps]{revtex}
\begin{document}
\draft
\title{\bf Parity and time reversal violating nuclear polarizability}
\author{V.V. Flambaum, J.S.M. Ginges, and G. Mititelu}
\address{School of Physics, University of New South Wales, 
Sydney 2052,Australia}
\date{\today}
\maketitle

\tightenlines

\begin{abstract} 

We propose a nuclear mechanism which can induce an atomic electric 
dipole moment (EDM). Parity and time reversal violating ($P,T$-odd) 
nuclear forces generate a mixed $P,T$-odd nuclear polarizability $\beta _{ik}$ 
(defined by an energy shift $U = -\beta _{ik} E_{i}H_{k}$, ${\bf E}$ is 
electric field and ${\bf H}$ magnetic field). 
The interaction of atomic electrons with $\beta _{ik}$ produces an 
atomic EDM. We performed an analytical calculation of the $P,T$-odd 
nuclear polarizability and estimated the value for the induced atomic EDM.   
The measurements of atomic EDMs can provide information about 
$P,T$-odd nuclear forces and test models of CP-violation.
 
\end{abstract} 
\vspace{1cm}
\section{Calculation of $P,T$-odd nuclear polarizability}
\label{nuclear}

Time ($T$) reversal violation in the interactions of elementary particles 
has only been indirectly observed in the $CP$-violating decay of the 
neutral K-mesons.
The observation of a $P,T$-odd multipole 
(for example, the dominant electric dipole moment (EDM)) of, for instance, 
an elementary particle or atom would be direct evidence for $T$-odd 
interactions. 
There are several schemes for incorporating $CP$-violation into unifying 
theories. These schemes make very different predictions for the value 
of EDMs of particles and atoms. While there has never been an unambiguous 
detection of these EDMs, upper limits on their values provide stringent 
tests of competing models of $CP$-violation.  

In this paper we propose a mechanism which can induce an atomic EDM: 
the interaction of a $P,T$-odd perturbed nucleus with
the electric and magnetic fields of an external electron.

The interaction of external electric ${\bf E}$ and magnetic ${\bf H}$ fields 
with operators of nuclear electric and magnetic dipole moments, 
${\bf d}$ and ${\bf \mu}$, 
gives rise to an energy shift, $U=-\beta _{ik}E_{i}H_{k}$, where 
$\beta _{ik}$ is the mixed $P,T$-odd nuclear polarizability 
\cite{Feinberg}.   
Parity and time reversal violating ($P,T$-odd) nuclear forces generate this 
polarizability,
\begin{equation}
\label{beta}
\beta _{ik}=
2\sum_{n}\frac{\langle \tilde{\psi}_0|d_{i}|\tilde{\psi}_n\rangle 
\langle \tilde{\psi}_n|\mu _{k}|\tilde{\psi}_0\rangle}
{E_{n}-E_{0}},
\end{equation}
where $\tilde{\psi}_{0}$ and $\tilde{\psi}_{n}$ are 
$P,T$-odd perturbed ground and excited nuclear states, 
respectively.
We consider a heavy spherical nucleus with a single unpaired nucleon.
The $P,T$-odd interaction between the 
non-relativistic nucleon and the nuclear core is given by (see, e.g. 
\cite{SFK1984})
\begin{equation}
\hat{H}_{PT}=
\frac{G}{\sqrt{2}}\frac{\eta}{2m}{\bf \sigma}{\bf \nabla}\rho ({\bf r}),    
\end{equation}
$G$ is the Fermi weak interaction constant, $\eta$ is a constant which 
characterizes the strength of the $P,T$-odd nucleon-nucleon interaction 
(limits on $\eta$ have been obtained from atomic \cite{atom95}
and molecular \cite{mol1989} experiments), 
${\bf \sigma}$ is the nuclear spin operator, 
$\rho ({\bf r})$ is the density of the nuclear core, 
and $m$ is the proton mass. 
The $P,T$-odd perturbed nuclear wavefunctions resulting from the interaction 
$\hat{H}_{PT}$, in an approximation where the nuclear density $\rho$ 
and potential $U$ coincide, $\rho _{0}(r)=U(r)\rho _{0}(0)/U(0)$, 
have the form \cite{SFK1984}
\begin{equation}
\tilde{\psi}_{n}=(1-\xi {\bf \sigma}\cdot {\bf \nabla})\psi _{n},
\end{equation}
where $\xi=-\eta \frac{G}{2\sqrt{2}m}\frac{\rho _{0}(0)}{U(0)}$ and  
$\psi _{n}$ is an unperturbed wavefunction. 

In lowest order in the $P,T$-odd interaction the polarizability 
(Eq. \ref{beta}) is
\begin{eqnarray}
\label{beta1}
\beta _{ik} 
&=&\sum_{n}\frac{2\xi}{E_{n}-E_{0}}
\{
\langle \psi _{0}|[{\bf \sigma}\cdot {\bf \nabla},d_{i}]|\psi _{n}\rangle 
\langle \psi _{n}|\mu _{k}|\psi _{0}\rangle +
\langle \psi _{0}|d_{i}|\psi _{n}\rangle 
\langle \psi _{n}|[{\bf \sigma}\cdot {\bf \nabla},\mu _{k}]|\psi _{0}\rangle 
\} \\
&=&{\beta _{ik}^{(1)}} + {\beta _{ik}^{(2)}}.
\end{eqnarray}
In the non-relativistic approximation, 
the only contribution to the first term of this
equation arises due to transitions between fine-structure doublets, 
for example $p_{1/2}$ and $p_{3/2}$, $d_{3/2}$ and $d_{5/2}$ 
(because the only non-zero matrix elements of ${\bf \mu}$ are between 
these states).
In the second term, states $\psi _{n}$ must have opposite parity to 
$\psi _{0}$ (because the matrix element of ${\bf d}=e{\bf r}$ 
cannot mix states of the same parity). 
The largest contributions to this term arise due to transitions
to the next shell. Therefore the energy denominator of the second 
term is much larger than that of the first.  

The nuclear electric dipole operator is defined as ${\bf d}=e_{N}{\bf r}$, 
where $e_{N}$ is the effective charge for the nucleon $N=n,p$, 
$e_{p}=(N/A)e$, $e_{n}=-(Z/A)e$, which appears due to the recoil effect.
The nuclear magnetic dipole operator is defined as 
${\bf \mu}=\mu _{N}(g_{l}{\bf l}+g_{s}{\bf s})$, where
$\mu _{N}$ is the nuclear magneton,
$g_{s}$ and $g_{l}$ are the spin and orbital $g$-factors; 
for the proton $g_{s}=5.586$, $g_{l}=1$ 
while for the neutron $g_{s}=-3.826$, $g_{l}=0$.

The first term of Eq. \ref{beta1} can be expressed as  
\begin{equation}
\beta _{ik}^{(1)}=
e_{N}\mu _{N}\xi (g_{s}-g_{l})\frac{1}{\omega _{\rm fs}}
\langle \psi _{0}|\sigma _{i}|\psi _{0'}\rangle 
\langle \psi _{0'}|\sigma _{k}|\psi _{0}\rangle ,
\end{equation}
where $\omega _{\rm fs}=E_{0'}-E_{0}$ is the energy difference between 
the fine structure components $\psi _{0}$ and $\psi _{0'}$.

Let us now consider the second term of Eq. \ref{beta1}, 
$\beta _{ik}^{(2)}$. This can be written as
\begin{equation}
\label{beta12}
\beta _{ik}^{(2)} = 
\sum_{n}\frac{\xi \mu _{N}}{E_{n}-E_{0}}
\{
g_{s}\langle \psi _{0}|d_{i}|\psi _{n}\rangle 
\langle \psi _{n}|[\sigma _{j}, \sigma _{k}]\nabla _{j}|\psi _{0}\rangle +
2g_{l}\langle \psi _{0}|d_{i}|\psi _{n}\rangle
\langle \psi _{n}|\sigma _{j}[\nabla _{j}, l_{k}]|\psi _{0}\rangle 
\}.
\end{equation}
Using the commutation relations
$[\sigma _{j}, \sigma _{k}]=2i\epsilon _{jkm}\sigma _{m}$ and 
$[\nabla _{j}, l_{k}]=i\epsilon_{jkm}\nabla _{m}$, we obtain
\begin{equation}
\label{beta12'}
\beta _{ik}^{(2)} = 
2i\xi \mu _{N}(g_{s}-g_{l})\epsilon_{jmk}
\sum_{n}\frac{\langle \psi _{0}|d_{i}|\psi _{n}\rangle
\langle \psi _{n}|\sigma _{j}\nabla _{m}|\psi _{0}\rangle }
{E_{n}-E_{0}} . 
\end{equation} 
This term can be further simplified by replacing the operator 
${\bf \nabla}$ by the commutation relation ${\bf \nabla}=m[{\bf r},\hat{H}]$,
where $\hat{H}=\hat{p}^{2}/2m +\hat{V}(r)$ is the single-particle 
Hamiltonian. Neglecting the spin-orbit interaction, that is setting 
$[\hat{H},{\bf \sigma}]=0$, and using closure, $\sum _{n}|\psi _{n}\rangle 
\langle \psi _{n}|=1$, Eq. \ref{beta12'} becomes
\begin{equation}
\beta _{ik}^{(2)} =  
-2i\xi e_{N}\mu _{N}m(g_{s}-g_{l})\epsilon _{kjm}
\langle \psi _{0}|r_{i}\sigma _{j}r_{m}|\psi _{0}\rangle .
\end{equation}

So, to first order in the $P,T$-odd interaction, we can write the 
nuclear polarizability as
\begin{equation}
\label{beta1fin}
\beta _{ik}=\xi e_{N}\mu _{N}(g_{s}-g_{l})
\{
(1/\omega _{\rm fs} )\langle \psi _{0}|\sigma _{i}|\psi _{0'}\rangle 
\langle \psi _{0'}|\sigma _{k}|\psi _{0}\rangle -
i2m\langle \psi _{0}|r_{i}({\bf \sigma}\times {\bf r})_{k}|\psi _{0}\rangle
\}.
\end{equation}

The $P,T$-odd nuclear polarizability can be expressed in terms of 
scalar and tensor components,
\begin{equation}
\label{betared}
\beta _{ik}=\delta _{ik}\beta _{s}+[I_{i}I_{k}+I_{k}I_{i}-\frac{2}{3}
\delta _{ik}I(I+1)]\beta _{t},
\end{equation}
$I$ is the nuclear total angular momentum.
We are interested in finding the scalar and tensor $P,T$-odd polarizabilities,
$\beta _{s}$ and $\beta _{t}$.
Let us start with the scalar contribution. Now, from Eq. \ref{betared}, 
we see that $\beta _{s}=\beta _{ii}/3$. We can see from Eq. \ref{beta1fin} 
that there is no scalar contribution from the second term, since 
$r_{i}({\bf \sigma}\times {\bf r})_{i}=0$. We can therefore write the 
scalar term as
\begin{equation}
\beta _{s}=\xi e_{N}\mu _{N}(g_{s}-g_{l})(\frac{1}{3\omega _{\rm fs}})
\langle \psi _{0}|\sigma _{i}|\psi _{0'}\rangle 
\langle \psi _{0'}|\sigma _{i}|\psi _{0}\rangle .
\end{equation} 
Evaluating this expression using spherical spinors, with 
the ground state wavefunction $\psi _{0}$ corresponding to $I=l+1/2$ 
and its fine-structure partner $\psi _{0'}$ corresponding to $I'=l-1/2$, 
$l$ is the orbital angular momentum, we obtain
\begin{equation}
\label{betas}
\beta _{s}=\xi e_{N}\mu _{N}(g_{s}-g_{l})(\frac{1}{\omega _{\rm fs}})
(\frac{2I-1}{3I}) .
\end{equation}

Let us now consider the tensor polarizability $\beta _{t}$. 
Setting $i,k=z$ in Eq. \ref{beta1fin} 
and taking the maximum projection, $I_{z}=I=l+1/2$, the first term  
is zero, since $\psi _{0'}$ with 
maximum projection $I'_{z}=I'=l-1/2$ gives zero contribution.
The second term is also zero in this case, 
since $\sigma _{x}$ and $\sigma _{y}$ cannot have diagonal matrix elements. 
Using Eq. \ref{betared} with $\beta _{zz}=0$ we obtain 
\begin{equation}
\label{betat}
\beta _{t}=-\frac{3\beta_{s}}{2I(2I-1)}.
\end{equation}

Inserting Eqs. \ref{betas} and \ref{betat} into Eq. \ref{betared}, we 
obtain for the nuclear $P,T$-odd polarizability
\begin{equation}
\label{betaik}
\beta _{ik}=\xi e_{N}\mu _{N}(g_{s}-g_{l})(\frac{1}{\omega _{\rm fs}})
(\frac{1}{I})
\{
\frac{(2I-1)}{3}\delta _{ik}-
\frac{1}{2I}[I_{i}I_{k}+I_{k}I_{i}-\frac{2}{3}\delta _{ik}I(I+1)]
\}.
\end{equation}
Note that in spherical nuclei the tensor part arises due to unpaired 
nucleons only. The scalar part does not require nuclear spin, 
therefore several nucleons can contribute to it; however, if all states  
in a fine-structure doublet are occupied, then the contributions 
cancel exactly.

It is interesting to consider the $P,T$-odd nuclear polarizability 
of a deformed nucleus. 
In this case we would expect contributions to the EDM from many nuclei 
(collective polarizability) because in deformed nuclei about $A^{2/3}$ 
nucleons are in open shells. This enhancement is similar to that 
of the $P,T$-odd nuclear quadrupole moment discussed in \cite{F1994}.
   
\section{Estimate for induced atomic EDM}
\label{atomic}

Consider an atom with one external electron above closed shells. 
We are interested in the atomic EDM induced by the interaction of
atomic electrons with the $P,T$-odd perturbed atomic nucleus polarized 
by the fields of the external electron (we use the nuclear model defined 
in the previous section).
The induced atomic EDM arises in the third order of perturbation theory,
\begin{equation}
d_{\rm atom}=\sum_{n,n',m,m'}
\frac{\langle 0,0'|d_{z}|n',n\rangle
\langle n,n'|\hat{H}_{E}|m',m\rangle 
\langle m, m' |\hat{H}_{B}|0',0\rangle }
{(E_{0}+E_{0'}-E_{n}-E_{n'})(E_{0}+E_{0'}-E_{m}-E_{m'})}
+ {\rm permutations},
\end{equation}
where unprimed states ($m,~n,~0$) denote electronic states, 
primed states ($m',~n',~0'$) denote ($P,T$-odd perturbed) nuclear states, 
and the electric and magnetic interactions between the  
external electron and the unpaired nucleon are
\begin{equation}
\label{H_EH_B}
\hat{H}_{E}=-\frac{e}{R^{3}}{\bf d}_{n}\cdot {\bf R}
\qquad 
\hat{H}_{B}=-\frac{e}{R^{3}}{\bf \mu }_{n}\cdot ({\bf R}\times {\bf \alpha}) ,
\end{equation}
${\bf d}_{n}$ and ${\bf \mu}_{n}$ are the electric and magnetic dipole moments 
of the nucleus, and ${\bf \alpha }$ is the electron Dirac matrix.
We use the Dirac interaction because the external electron close to the 
nucleus is relativistic.

To measure an EDM, an external electric field is applied to the system 
which couples to the electron electric dipole moment operator $d_z$. 
Terms in which this operator admixes excited states, 
for example
\begin{equation}
\sum_{m,n,n'}\frac{\langle 0,0'|\hat{H}_{E}|n',n\rangle
\langle n,n'|d_{z}|n',m\rangle 
\langle m, n' |\hat{H}_{B}|0',0\rangle }
{(E_{0}+E_{0'}-E_{n}-E_{n'})(E_{0}+E_{0'}-E_{m}-E_{n'})} \nonumber
\end{equation}
have very large (nuclear) energy denominators as here the 
electric field interacts with the excited states of the atom.
We can therefore neglect this contribution to the atomic EDM.

The atomic EDM induced by the interactions (Eq. \ref{H_EH_B}) can therefore 
be presented as
\begin{equation}
\label{EDM}
d_{\rm atom}\approx 2\sum _{n}\frac{\langle 0|d_{z}|n\rangle}
{(E_{0}-E_{n})} 
[ 2\sum_{m'}\frac{\langle n,0'|\hat{H}_{E}|m'\rangle 
\langle m'|\hat{H}_{B}|0',0\rangle}{(E_{0'}-E_{m'})} ],
\end{equation}
where we have set $n'=0'$, because $d_{z}$ does not act on the nuclear 
wavefunctions.
Typical electron energies are much smaller than nuclear energies,
$(E_{0}-E_{m})<<(E_{0'}-E_{m'})$, therefore we have used closure for 
the electron states, $\sum_{m}|m\rangle \langle m|=1$.
Substituting Eq. \ref{H_EH_B} into Eq. \ref{EDM}, and using the 
definition for the $P,T$-odd nuclear polarizability (Eq. \ref{beta}),
we obtain
\begin{equation}
\label{EDMbeta}
d_{\rm atom}\approx
2\frac{e^{2}}{R^{6}}
\epsilon _{klp}\beta _{ik}
\sum _{n}\frac{\langle 0|d_{z}|n\rangle 
\langle n|R_{i}R_{l}\alpha _{p}|0\rangle}
{E_{0}-E_{n}}
\equiv
2\sum _{n}\frac{\langle 0|d_{z}|n\rangle 
\langle n|\hat{H}_{PT}|0\rangle}
{E_{0}-E_{n}} ,
\end{equation}
where we have defined an effective Hamiltonian $\hat{H}_{PT}$ which 
describes the interaction of the $P,T$-odd nuclear polarizability with 
atomic electrons. 

In this approximation, the contribution of the scalar polarizability 
to the atomic EDM is zero (${\bf R}\cdot [{\bf R}\times {\bf \alpha}]=0$).
However, there is no theorem which forbids this contribution. 
Indeed, consider non-relativistic expressions for magnetic and electric 
fields produced by a charged particle with magnetic moment 
${\bf \mu}=\mu {\bf s}$,
\begin{equation}
{\bf H}=e\frac{{\bf n}\times {\bf v}}{R^{2}}+
\frac{3({\bf n}\cdot {\bf \mu}){\bf n}-{\bf \mu}}{R^{3}}+
\frac{8\pi}{3}{\bf \mu}\delta (R)  
\qquad
{\bf E}=e\frac{\bf n}{R^{2}},
\end{equation}
where ${\bf v}$ is the velocity of the particle, ${\bf n}={\bf R}/R$.
Substituting these expressions into the interaction energy
\begin{equation}
\label{nonrel}
\beta _{s}\delta _{ik}H_{i}E_{k}=2e\mu \beta _{s}
\frac{{\bf s}\cdot {\bf n}}{R^{2}}[\frac{1}{R^{3}}+\frac{4\pi}{3}
\delta (R)]
\end{equation}
we see that the first, longer-range orbital field in the expression for 
${\bf H}$ indeed does not give any contribution to the interaction.
However, the spin magnetic field, which decays more rapidly with distance, 
does contribute.
The angular structure of the effective interaction (Eq. \ref{nonrel}) 
is similar to that between an electron EDM and an atomic electric field 
(see, e.g., \cite{Khripbook}).
However, the interaction (Eq. \ref{nonrel}) decays with distance very 
rapidly, and is very singular at small distances. 
This strong singularity is an indication that one should use 
the relativistic form for the magnetic interaction (Eq. \ref{H_EH_B}) 
in order to describe the small distance contribution correctly. 
We showed above that in this case our calculation gives zero 
for the $\beta _{s}$ contribution to the EDM.

The only contribution of the nuclear $P,T$-odd polarizability 
to the atomic EDM therefore arises due to the tensor term.
Inserting the expression for $\beta _{ik}$ (Eq. \ref{betaik}) 
into the above expression (Eq. \ref{EDMbeta})
we obtain for the effective Hamiltonian $\hat{H}_{PT}$
\begin{equation}
\label{HPT}
\hat{H}_{PT}= -\frac{e^{2}}{2R^{6}}e_{N}\xi \mu _{N}(g_{s}-g_{l})
(\frac{1}{\omega _{\rm fs}})(\frac{1}{I})^{2}\epsilon _{klp}
R_{i}R_{l}\alpha _{p}
[I_{i}I_{k}+I_{k}I_{i}-\frac{2}{3}\delta _{ik}I(I+1)].
\end{equation}
In earlier works the value of atomic EDMs of Cs \cite{K1976,SFK1984}, 
Dy \cite{DFK1986,F1994} and Ra \cite{F1999,DFG2000} induced by the 
($P,T$-odd) nuclear magnetic quadrupole moment (MQM) have been calculated.
It is therefore useful to compare
$\hat{H}_{PT}$ with the Hamiltonian $\hat{H}_{MQM}$ which describes 
the interaction of a nuclear MQM with 
atomic electrons,
\begin{equation}
\label{MQM}
\hat{H}_{MQM}
=\frac{1}{R^{5}}\frac{Me}{2I(2I-1)}
\epsilon _{klp}R_{i}R_{l}\alpha _{p}
\{ I_{i}I_{k}+I_{k}I_{i}-\frac{2}{3}\delta _{ik}I(I+1) \},
\end{equation}
in order to obtain an estimate for the value of the atomic 
EDM induced by the nuclear polarizability.
Here $M$ is the magnetic quadrupole moment which, for $I=l+1/2$, is 
$M=-\xi \mu _{N}(g_{s} - 2g_{l})(2I-1)$. 
The ratio of $\hat{H}_{PT}$ to $\hat{H}_{MQM}$ is
\begin{equation}
\frac{\hat{H}_{PT}}{\hat{H}_{MQM}}
=\frac{1}{R}\frac{(g_{s}-g_{l})ee_{N}}{(g_{s}-2g_{l})\omega _{\rm fs}I}.
\end{equation}

The largest contribution to the atomic EDM arises due to mixing between 
$s$ and $p_{3/2}$ electron states. 
The radial integral for the magnetic quadrupole converges at 
$R\sim a_{B}/Z$ \cite{SFK1984}, where $a_{B}$ is the Bohr radius. 
The $\hat{H}_{PT}$ matrix element 
is more singular at the nucleus than that of $\hat{H}_{MQM}$. 
Considering $s$ and $p_{3/2}$ mixing, the EDM induced by the 
nuclear polarizability contains an extra factor
\begin{equation}
S_{\rm rel}\equiv 
-\frac{\gamma _{1}+\gamma _{2}-2}{\gamma _{1}+\gamma _{2}-3}
(\frac{a_{B}}{ZR_{N}})^{3-\gamma _{1}-\gamma _{2}},
\end{equation}
where $R_{N}$ is the nuclear radius,
$\gamma _{1}=\sqrt{1-Z^{2}\alpha ^{2}}$ and 
$\gamma _{2}=\sqrt{4-Z^{2}\alpha ^{2}}$.
For $Z^{2}\alpha ^{2}<<1$,
\begin{equation}
S_{\rm rel}\approx \frac{4}{3}\frac{1}{Z^{2}\alpha ^{2}}
(\frac{ZR_{N}}{a_{B}})^{-\frac{3}{4}Z^{2}\alpha ^{2}}
\end{equation} 
has a very weak dependence on the nuclear radius $R_{N}$ which we use as a 
cut-off parameter in the integration over $R$.
Therefore, the ratio of $\hat{H}_{PT}$ to $\hat{H}_{MQM}$ is 
\begin{equation}
\frac{\hat{H}_{PT}}{\hat{H}_{MQM}}\sim
\frac{Ze^{2}S_{\rm rel}}{a_{B}\omega _{\rm fs}}.
\end{equation}
For heavy atoms this ratio is of the order of 1\%. 

We are grateful to O.P. Sushkov and M.Yu. Kuchiev for useful discussions.
This work was supported by the Australian Research Council.


\end{document}